\documentclass[12pt]{iopart}
\usepackage{graphicx}

\begin{document}

\title{Role of noise and agents' convictions on opinion spreading in a three-state voter-like model}

\author{Nuno Crokidakis}
\address{
Departamento de F\'isica, PUC-Rio, Rio de Janeiro, Brazil
}

\ead{nuno.crokidakis@fis.puc-rio.br}

\begin{abstract}
\noindent
In this work we study the opinion formation in a voter-like model defined on a square lattice of linear size $L$. The agents may be in three different states, representing any public debate with three choices (yes, no, undecided). We consider heterogeneous agents that have different convictions about their opinions. These convictions limit the capacity of persuasion of the individuals during the interactions. Moreover, there is a noise $p$ that represents the probability of an individual spontaneously change his opinion to the undecided state. Our simulations suggest that the system reaches stationary states for all values of $p$, with consensus states occurring only for the noiseless case $p=0$. In this case, the relaxation times are distributed according to a log-normal function, with the average value $\tau$ growing with the lattice size as $\tau\sim L^{\alpha}$, where $\alpha\approx$ 0.9. We found a threshold value $p^{*}\approx$ 0.9 above which the stationary fraction of undecided agents is greater than the fraction of decided ones. We also study the consequences of the presence of external effects in the system, which models the influence of mass media on opinion formation.

\end{abstract}

\maketitle


\section{Introduction}

\qquad Models of opinion formation have been studied by physicists since the 80's and are now part of the new branch of physics called sociophysics \cite{galam_book}. This recent research area uses tools and concepts of the physics of disordered matter and more recently from the network science to describe some aspects of social and political behavior \cite{galam_review,loreto_rmp}. From the theoretical point of view, opinion models are interesting to physicists because they present correlations, order-disorder transitions, scaling and universality, among other typical features of physical systems \cite{loreto_rmp}.

One of the most studied models to analyze the dynamics of agreement and disagreement is the voter model (VM) \cite{loreto_rmp}. The VM was introduced to study the competition of species \cite{clifford} and was firstly named voter model in the work of Holley and Ligget \cite{holley}. It represents a very simple formulation of opinion dynamics, where each agent carries one opinion given by an Ising variable $s=\pm 1$ and a randomly choosen agent takes the opinion of one of his neighbors at each time step. In other words, the VM is a paradigmatic example of nonequilibrium copying dynamics, where the agents imitate their neighbors. Concerning regular lattices, this model can be exactly solved in any dimension \cite{redner,krapivsky}. Besides the simplicity of the model, it was shown that the two-dimensional VM represents a broad class of models, defining a kind of voter-model universality class \cite{dornic}. 

The VM was extended to take into account different ingredients. As examples, we can cite the generalization to three \cite{vazquez} or more variables \cite{sire}, the inclusion of special agents like zealots \cite{mobilia}, introduction of memory \cite{castellano} and time-dependent transition rates \cite{stark}, and others. The model was also studied in several types of complex networks, that have nontrivial effects on the ordering dynamics (see \cite{loreto_rmp} and references therein).

In addition to the interactions among individuals, it is also interesting from theoretical and practical points of view to analyze the influence of external effects on agent-based social models \cite{moscovici,galam_rfim}. In opinion dynamics, for example, these external effects act in the system as a mass media (television, radio, ...). Among the previous studies considering mass-media effects, one can cite the Sznajd opinion model \cite{sznajd_media,schulze,meu_physicaA}, the Axelrod model of culture diversity \cite{avella,rodriguez,candia,peres} and some other social models \cite{pabjan,candia2,guo,babak,sirbu}. The external effects make social models more realistic, and they may produce interesting results like the induction/suppression of phase transitions, the decrease of the relaxation times and the emergence/lack of consensus.

In this work we study a three-state opinion model, where the agents' states or opinions are represented by variables $s=+1$, $-1$ or $0$. The agents change their opinions via two competing mechanisms. Two agents may interact via a voter-model dynamics, but the usual imitation process is limited by a particular feature of each agent, his conviction, a characteristic that was considered in some opinion models \cite{ben-naim1,ben-naim2,ben-naim3,sen,souza,meu_celia}. In addition, there is a noise in the system that allows the agents to change their opinions to $s=0$. We analyze the consequences of these two ingredients, convictions and noise, on opinion formation and on opinion spreading across the population. After the initial analysis, we add an external effect that acts in the system as a mass medium, and we study the consequences of the presence of this external effect on the behavior of the model.

This work is organized as follows. In Section 2 we present the miscroscopic rules that define the model. The numerical and analytical results are discussed in Section 3, and our conclusions are presented in Section 4.


\section{Model}

\qquad We have considered a square lattice of size $L\times L$ with periodic boundary conditions as a simple representation of the social contacts among individuals. Each agent $i$ in the population may be in one of three possible states, namely $s_{i}=+1$, $-1$ or $0$. This scenario can represent any public debate, for example an electoral process with 2 different candidates A and B where each agent (or elector) votes for the candidate A (opinion $+1$), for the candidate B (opinion $-1$) or is undecided (opinion $0$). In addition, each agent has a conviction $C_{i}$ about his opinion. In the case of undecided voters, this conviction is given by $C=0$, and for the cases of opinions $+1$ and $-1$ each conviction is given by a random number generated from a uniform distribution $[0,1]$. The following microscopic rules control our model:

\begin{itemize}

\item We choose a random agent $i$;

\item We choose at random one of his nearest neighbors, say $j$; 

\item If $s_{i}=s_{j}$, nothing occurs;

\item On the other hand, if $s_{i}=-s_{j}$, the agent with the lower conviction, say $j$, is persuaded by $i$ and adopt both the opinion and the conviction of $i$. In other words, if $C_{i}>C_{j}$ we update $s_{j}=s_{i}$ and $C_{j}=C_{i}$;

\item If one of the two agents, say $j$, is undecided ($s_{j}=0$), he follows the opinion of agent $i$ and take his conviction about that opinion, i.e., we update $s_{j}=s_{i}$ and $C_{j}=C_{i}$.

\item We choose at random another nearest neighbor of $i$ (different from $j$), say $k$. With probability $p$ this agent $k$ becomes undecided, i.e., we update $s_{k}=0$ and $C_{k}=0$; 

\end{itemize}

Thus, the interactions are of voter-model type \cite{loreto_rmp}, but the usual imitation or copying process is limited by the agents' convictions. In addition, we considered that a certain persuaded agent takes not only the opinion but also the conviction of the agent that interacted with him. In the case of undecided voters, we considered that the convictions are $C=0$, and thus they are easily persuaded by decided voters \footnote{We have verified that the consideration of convictions $C\neq 0$ for undecided voters does not change qualitatively the results of the model, as discussed in the following.}. Notice also that the probability $p$ acts as a noise in the system, and it allows the spontaneous change of opinions (to $s=0$) of some agents \cite{ben-naim3}. Observe also the presence of intrinsic correlations in the system: the interacting agents $i$ and $j$, as well as the agent $k$ that suffers the effect of noise are always neighbors in the square lattice. Then, our dynamics inserts correlation effects on the noise itself. Some consequences of these correlations are discussed in the next section. In the following we will see that the presence of such heterogeneities in the population (convictions and noise) affects the persuasion process and the consensus states are obtained only in the absense of noise. We will also analyze an extension of the model considering an external effect, that acts in the system as a mass medium.

For the above-mentioned rules, we made the following assumptions: (i) as the agents interact by pairs, we have considered that if a given agent persuades another agent to change opinion (explaining the reasons for choosing a certain candidate, for example), the persuaded individual will became a supporter of the choosen opinion (candidate) and he will influence other individuals based on the ``learned'' reasons; (ii) undecided individuals or voters are passive, in the sense that they do not spread their lack of opinion to other individuals \cite{fontoura}; (iii) undecided agents are easily persuaded by interaction with someone that already has a formed opinion (which explain the null convictions); (iv) the noise $p$ represents the fact that some individuals in an electoral process choose a certain candidate at some time but they are not sure about their choices (in this case, those individuals may become undecided until interact with other agents).


\section{Results}

\subsection{The model in the absence of external effects}

\qquad In the absence of external effects, the model follows exactly the rules presented in the previous section. The initial state of the system is fully disordered, i.e., all opinions are equally probable ($1/3$ for each one). One time step in the model is defined by the application of the above-mentioned rules $N$ times, where $N=L^{2}$ is the total number of agents in the lattice. As a first analysis, we have considered the order parameter $O$ given by
\begin{equation}\label{eq1}
O = \frac{1}{N}\left|\sum_{i=1}^{N}s_{i}\right| ~,
\end{equation}
\begin{figure}[t]
\begin{center}
\vspace{0.5cm}
\includegraphics[width=0.4\textwidth,angle=0]{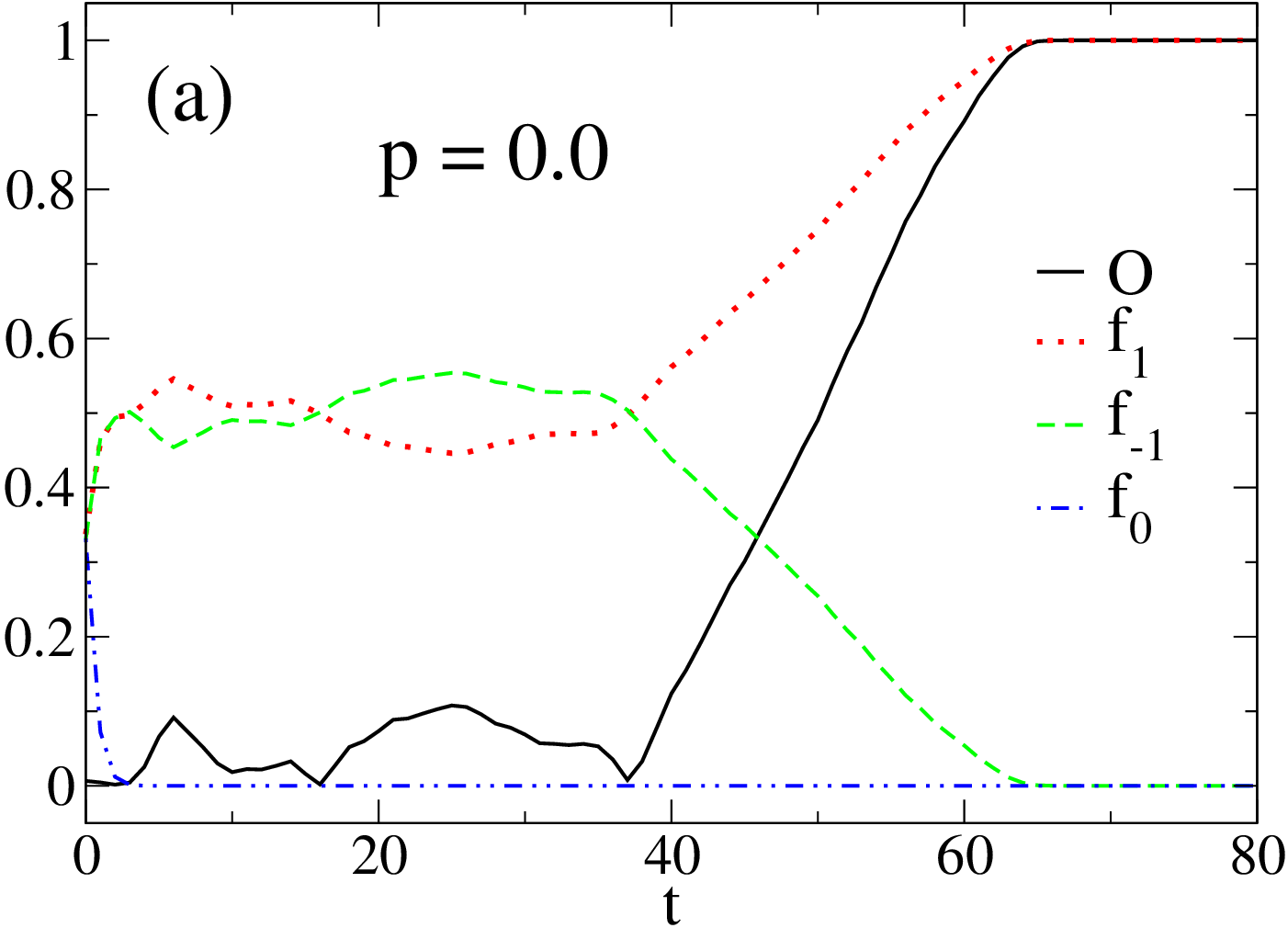}
\hspace{0.5cm}
\includegraphics[width=0.4\textwidth,angle=0]{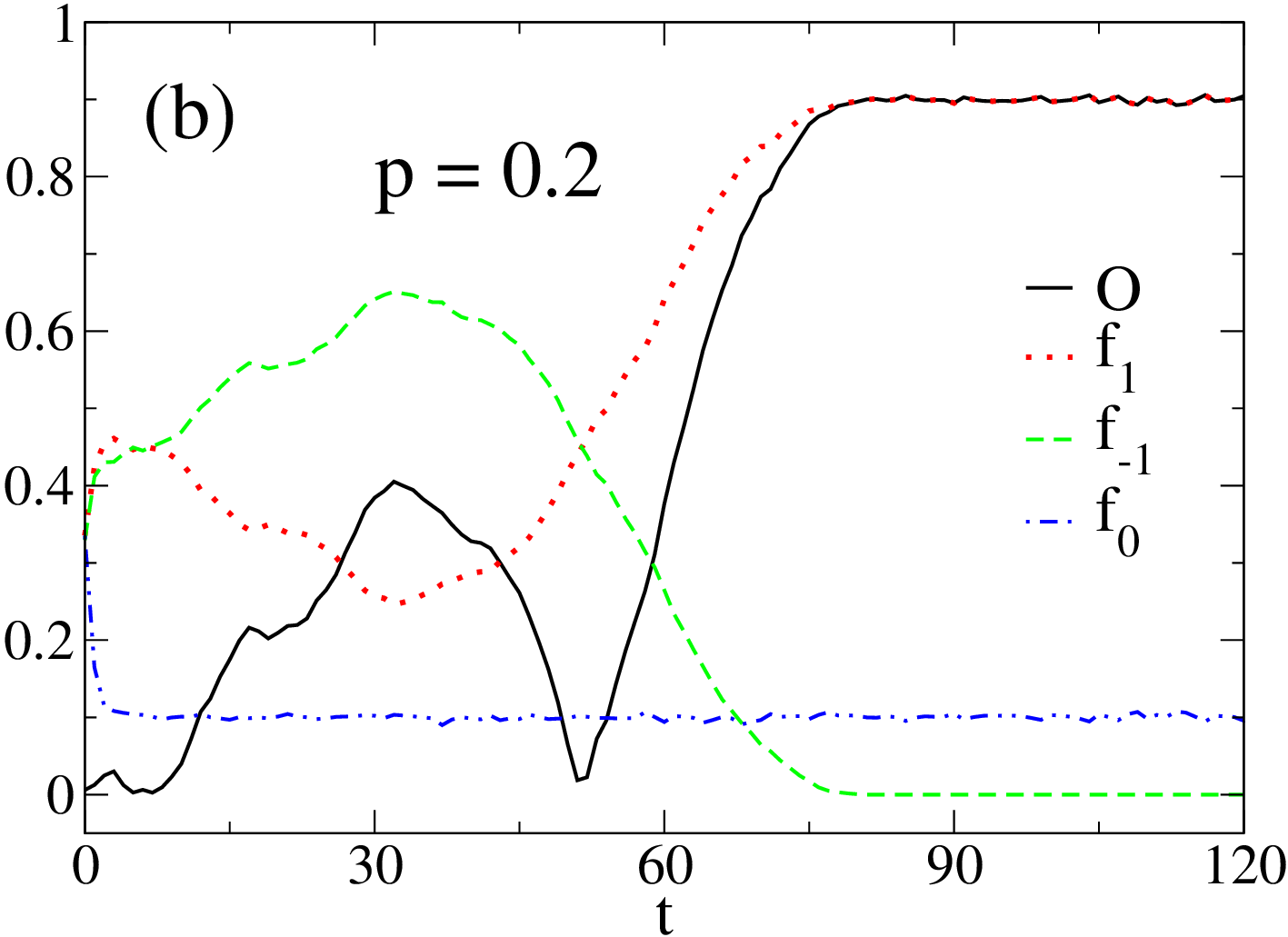}
\\
\vspace{1.0cm}
\includegraphics[width=0.4\textwidth,angle=0]{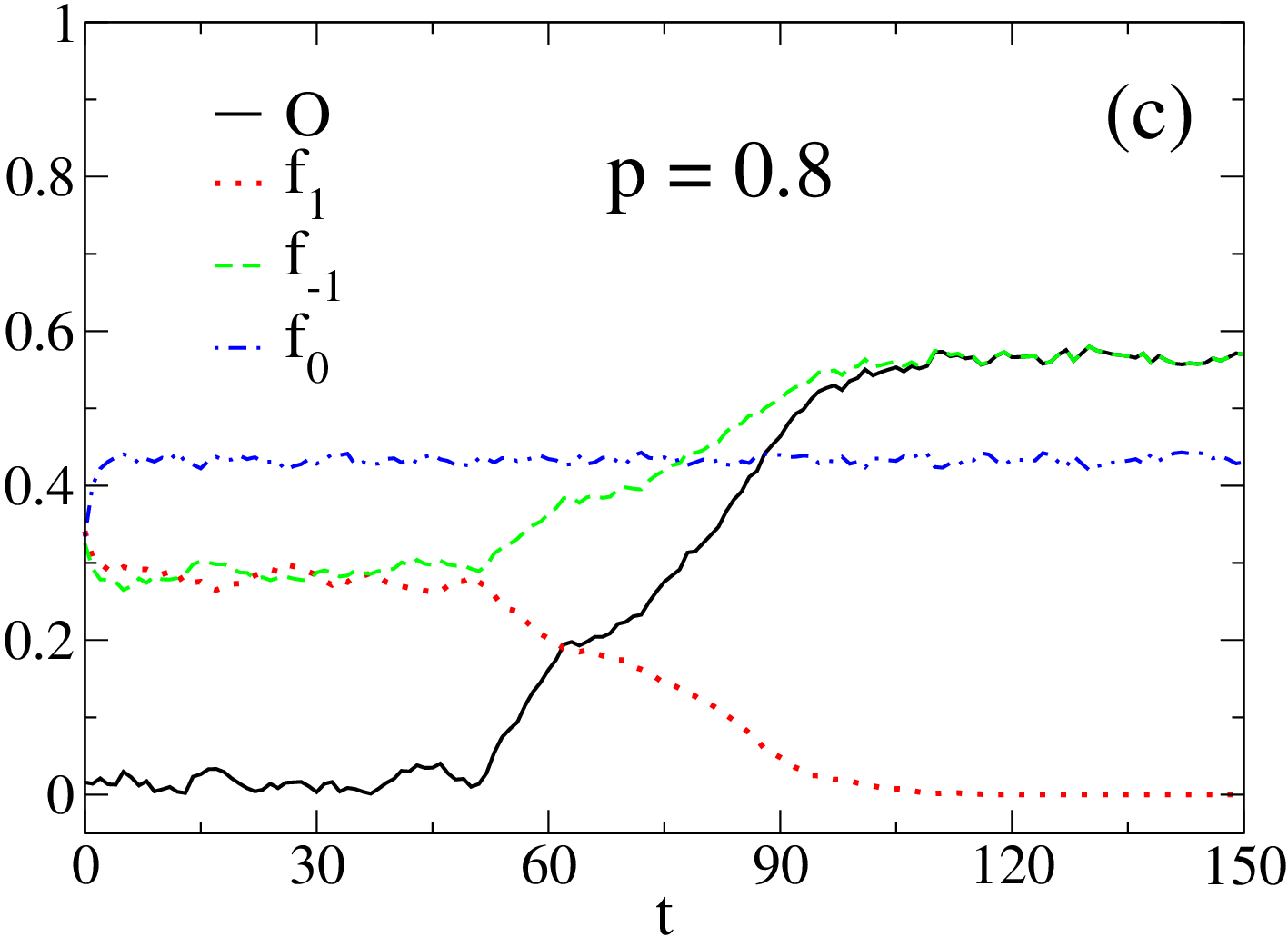}
\hspace{0.5cm}
\includegraphics[width=0.4\textwidth,angle=0]{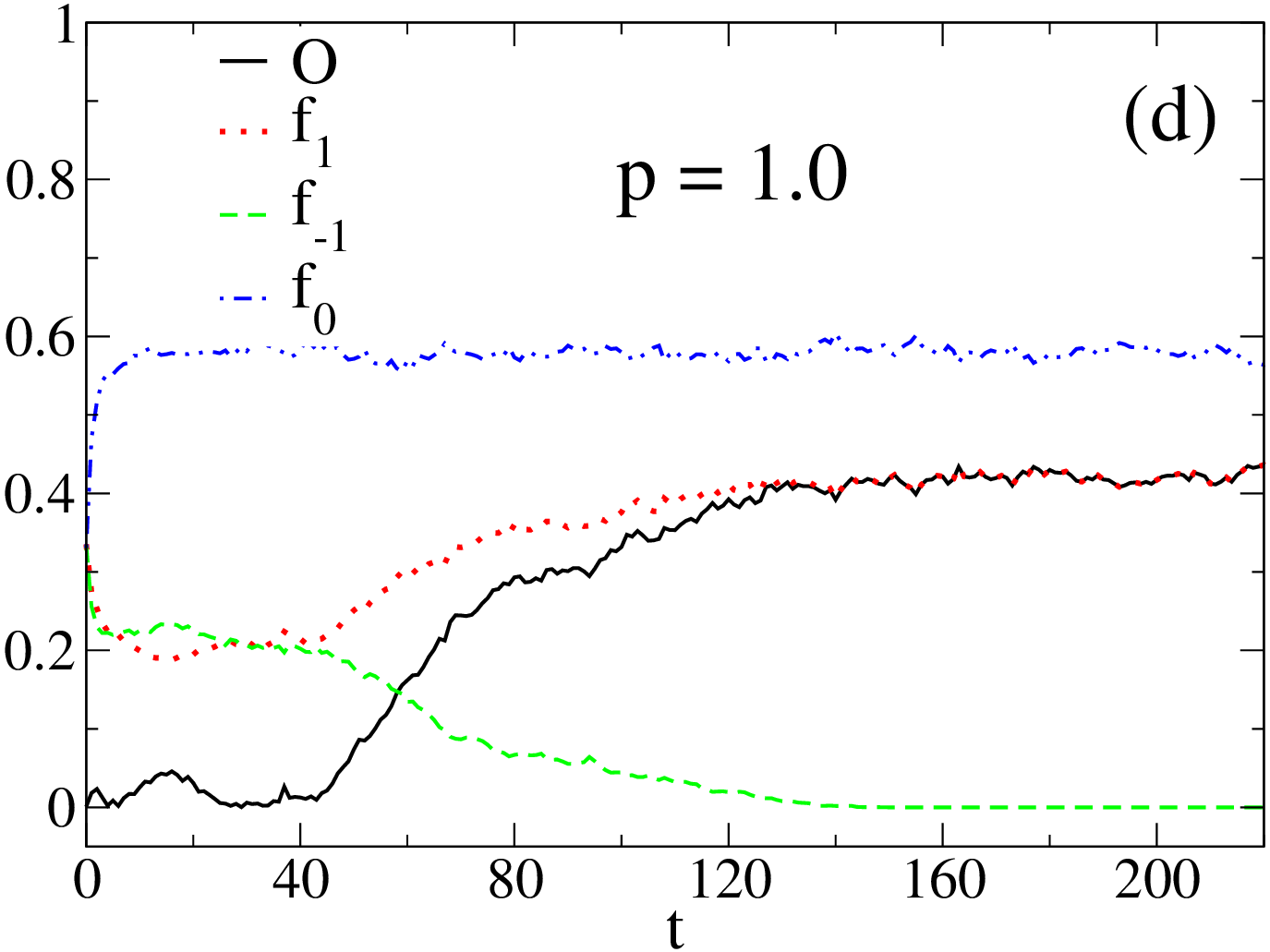}
\end{center}
\caption{Time evolution of the order parameter $O$ and the fractions $f_{1}$, $f_{-1}$ and $f_{0}$ of the agents with opinions $+1$, $-1$ and $0$, respectively, in the absence of external effects. Each graphic is a single realization of the dynamics on a square lattice of size $L=100$ and noise parameter $p=0.0$ (a), $p=0.2$ (b), $p=0.8$ (c) and $p=1.0$ (d).
}
\label{fig1}
\end{figure}
\noindent
that is the ``magnetization per spin'' of the system. In Fig. \ref{fig1} we exhibit the time evolution of the order parameter for typical values of the noise $p$. For all cases we have observed that in the steady states the fraction of one of the extreme opinions $+1$ or $-1$ are not found anymore in the population. Thus, in the stationary states the magnetization is given by $O=f_{1}$ ($O=f_{-1}$) in the case where the opinion $-1$ ($+1$) disappears of the system, where $f_{1}$ ($f_{-1}$) is the fraction of $+1$ ($-1$) opinions.  In Fig. \ref{fig1} we can see that the fraction $f_{0}$ of $s=0$ opinions (undecided individuals) for long times is different from zero, except for the noiseless case $p=0.0$ where we have $f_{0}=0$ in the steady states. We can also see from Fig. \ref{fig1} that the system reaches consensus with all opinions $+1$ or $-1$ only for $p=0.0$. These two results can be predicted analytically, as we will see in the following.

As discussed before, the initial state of the population is disordered such that the initial fraction of $s=0$ opinions is $f_{0}\approx 1/3$. We have verified numerically that the stationary fraction $f_{0}$ is greater than that initial value $1/3$ only for $p>\sim 0.7$, as it is shown in Figs. \ref{fig1} (c) and (d). We can also see from Fig. \ref{fig1} that the system needs more time to reach the steady states when we increase the noise $p$. In fact, for increasing values of $p$ more agents will change their opinions to $s=0$ such that more interactions among agents will occur, as well as more competition among the opinions. Another important result is that this final fraction $f_{0}$ of $s=0$ opinions is greater than the other surviving fraction ($f_{1}$ or $f_{-1}$) for sufficient large values of $p$. An example of this behavior is exhibited in Fig. \ref{fig1} (d), for the case $p=1.0$. We will discuss this in more details in the following.

\begin{figure}[t]
\begin{center}
\vspace{0.5cm}
\includegraphics[width=0.48\textwidth,angle=0]{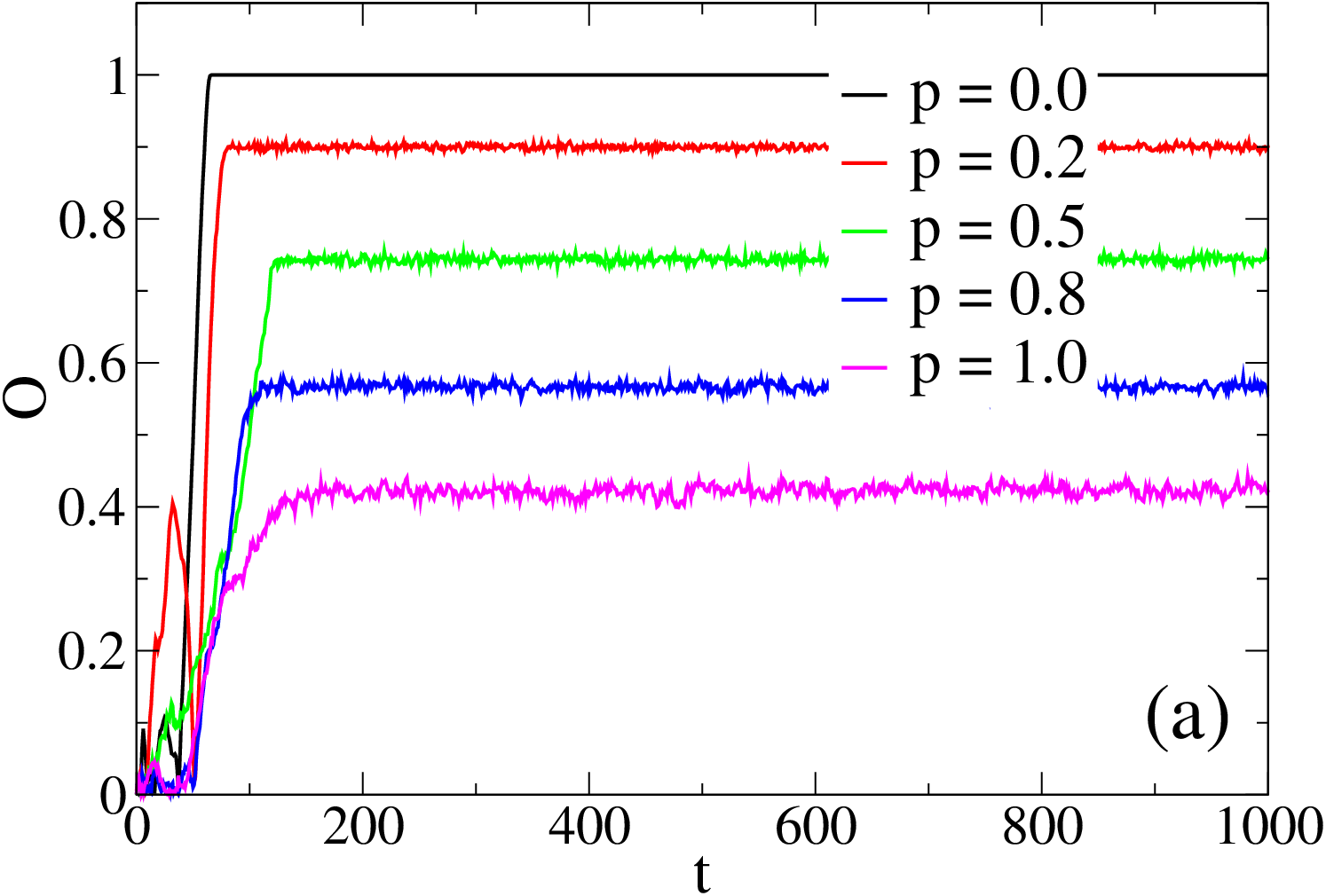}
\hspace{0.3cm}
\includegraphics[width=0.44\textwidth,angle=0]{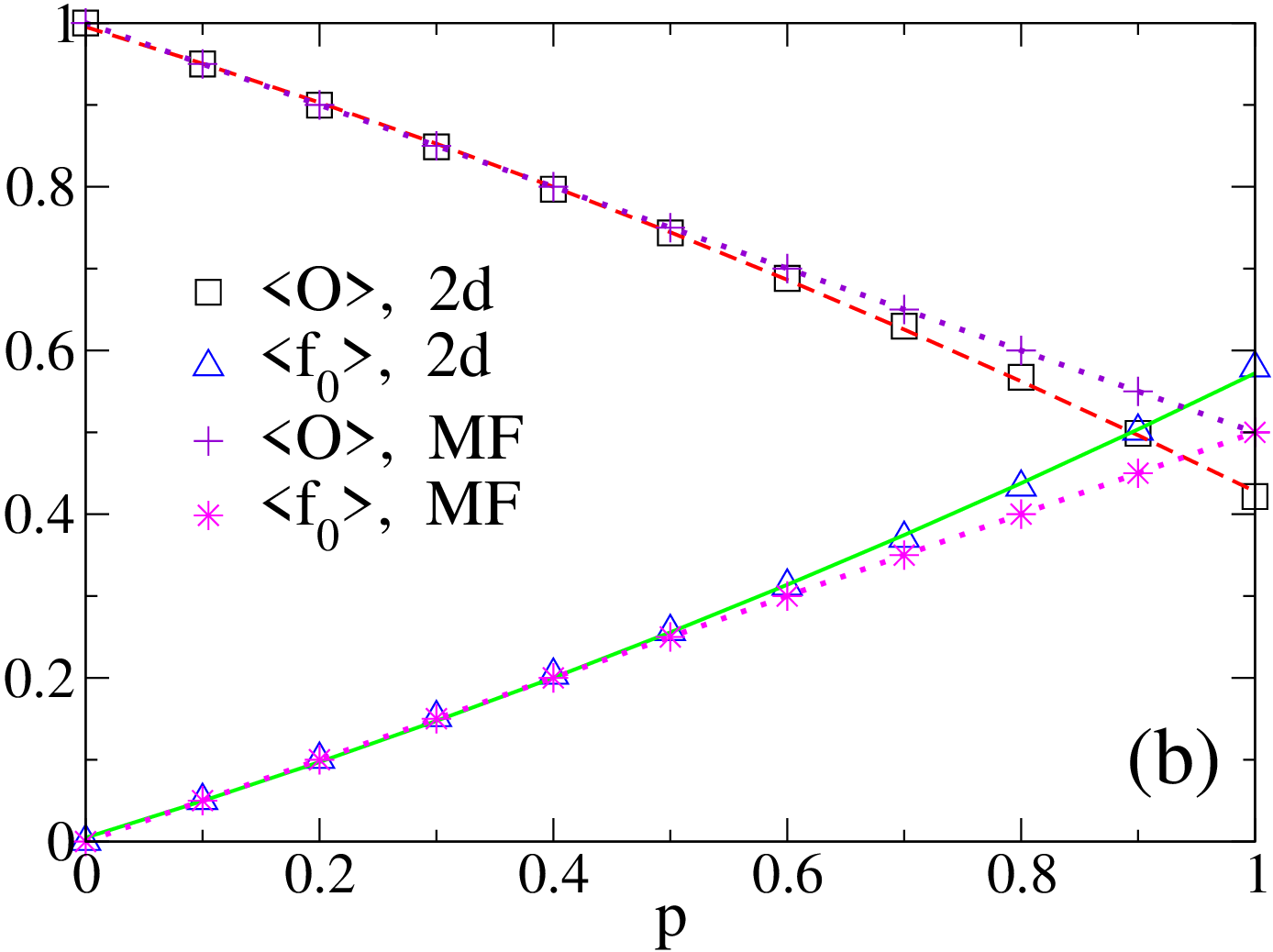}
\end{center}
\caption{Time evolution of the order parameter $O$ for typical values of the noise $p$. Each curve represents a single realization of the dynamics on a square lattice of size $L=100$ (a). It is also exhibited the average stationary values of the order parameter $\langle O \rangle$ and the fraction of undecided voters $\langle f_{0}\rangle$ versus $p$ obtained numerically for the 2d case and for the mean-field (MF) case for populations of size $N=10^{4}$ (b). Each symbol is averaged over $200$ independent simulations. The full and dashed lines are fits with second-order polynomials for the 2d case, whereas the dotted lines are the Eqs. (\ref{eq4}) and (\ref{eq5}), as explained in the text.
}
\label{fig2}
\end{figure}

We can study the steady-state properties of the model in more details. It is shown in Fig. \ref{fig2} (a) the time evolution of the order parameter $O$ for some values of the noise $p$, for a lattice size $L=100$. One can see that the stationary values of the magnetization decrease for increasing values of $p$, as observed before (see Fig. \ref{fig1}). We have considered the time averages of the stationary values of the order parameter $O$ for $200$ independent simulations, which give us the mean values $\langle O\rangle$ exhibited in Fig. \ref{fig2} (b). We can observe that there is no order-disorder transition: the lower value of $\langle O\rangle$ is about $0.42$. We also plotted in Fig. \ref{fig2} (b) the average value of the fraction of undecided voters $\langle f_{0}\rangle$ (also over $200$ realizations). As expected, when we increase the noise $p$ the fraction of agents with opinion $s=0$ increases. As discussed above, for sufficient large values of $p$ the final fraction $f_{0}$ is greater than the other surviving fraction ($f_{1}$ or $f_{-1}$). We can clearly see this behavior in Fig. \ref{fig2} (b): there is a threshold value $p^{*}$ above which the average stationary fraction $\langle f_{0}\rangle$ is greater than $\langle O\rangle$. To estimate the numerical value of $p^{*}$ we fitted the data for $\langle O\rangle$ and $\langle f_{0}\rangle$ with second-order polynomials \footnote{The simplest choice for a fitting function is a linear one, but it fails to fit the data. The choice for a parabolic function was done as a second attempt.} [see the dashed and full lines in Fig. \ref{fig2} (b)]. In this case, we can equate the polynomials to find
\begin{equation} \label{eq2}
p^{*} \approx 0.9  ~.
\end{equation}
\noindent
Above this threshold value $p^{*}$ there are more undecided individuals than decided ones in the population in the stationary states.

We can better understand the above-discussed behavior by deriving some equations at a mean-field (MF) level, following Refs. \cite{meu_celia,biswas}. The fractions $f_{1}$, $f_{-1}$ and $f_{0}$ of opinions $+1$, $-1$ and $0$, respectively, represent the probabilities of randomly choose an agent with opinion $s=+1$, $-1$ or $0$, respectively. One can consider the processes contributing to the in/out flux for $f_{0}$. Thus, the flux into $f_{0}$ is given by $p\,f_{1} + p\,f_{-1}=p\,(f_{1}+f_{-1})$, which corresponds to the sixth rule of Section 2 (the noise effect, i.e., the spontaneous change to opinion $s=0$). On the other hand, the flux out of $f_{0}$ is given by $2\,f_{0}\,(1-f_{0})$, which corresponds to pick randomly an agent $i$ with opinion $s_{i}=0$ (probability $f_{0}$) and then another random agent $j$ with opinion $s_{j}=+1$ or $-1$ (complementary probability $1-f_{0}$). The factor 2 in the last expression stands for the inverse case ($s_{i}=+1$ or $-1$ and $s_{j}=0$). In other words, this corresponds to the fifth rule of Section 2 (interactions among decided and undecided individuals). Considering the normalization condition, $f_{1}+f_{-1}+f_{0}=1$, one obtains $f_{1}+f_{-1}=1-f_{0}$. Taking into account that in the stationary state the fluxes into and out $f_{0}$ should be equal \cite{meu_celia,biswas}, the previous results lead to
\begin{equation} \label{eq3}
p\,(1-f_{0}) = 2\,f_{0}\,(1-f_{0}) ~. 
\end{equation} 
\noindent
There are two possible solutions (equilibrium points) for Eq. (\ref{eq3}), $f_{0}=1$ and $f_{0}=p/2$. A simple stability analysis of these solutions shows that the equilibrium point $f_{0}=1$ is unstable, so it can be neglected \footnote{We can also verify numerically that at least one of the fractions $f_{1}$ or $f_{-1}$ is $>0$, which implies that $f_{0}<1$ (not shown).}. On the other hand, the another equilibrium point $f_{0}=p/2$ is stable, which implies that the mean-field solution for the stationary fraction of undecided individuals is given by
\begin{equation} \label{eq4}
f_{0} = \frac{1}{2}\,p ~.
\end{equation} 
The order parameter is given by $O=|f_{1}- f_{-1}|$ which is equal to $O=|2\,f_{1}+f_{0}-1|$ by using the normalization condition. As suggested by the simulations, the stationary fraction of one of the extreme opinions $+1$ or $-1$ is equal to zero. In this case, let us say that we have $f_{-1}=0$ at the stationary state, which give us $O=f_{1}$ and
\begin{equation} \label{eq5}
O = 1 - f_{0} = 1-\frac{1}{2}\,p ~.
\end{equation} 
\noindent
Notice that the above Eqs. (\ref{eq4}) and (\ref{eq5}) predict three important behaviors: (i) there is no order-disorder transition, since the lower value of the order parameter is $O=1/2$ for $p=1.0$; (ii) the consensus states $O=1$ are obtained only for $p=0.0$, and (iii) the stationary fraction $f_{0}$ of undecided agents is equal to zero only for $p=0.0$. These results are in agreement with the simulations of the model on the square lattice.

We can confront these analytical calculations with numerical simulations for the model in the mean-field limit. In this case, all the three involved individuals ($i$, $j$ and $k$ in the notation of the rules in Section 2) are randomly choosen. In order to obtain the averages $\langle O\rangle$ and $\langle f_{0}\rangle$, we have performed $200$ independent simulations of populations of size $N=10^{4}$ agents. The result is exhibited in Fig. \ref{fig2} (b). One can see that the numerical results in the fully-connected lattice agree with the derived Eqs. (\ref{eq4}) and (\ref{eq5}), i.e., the stationary fraction $\langle f_{0}\rangle$ of undecided individuals grows linearly with the noise $p$, whereas the order parameter $\langle O\rangle$ decays linearly with $p$ [see the dotted lines in Fig. \ref{fig2} (b)]. Notice also from  Fig. \ref{fig2} (b) that there is no crossing of the curves of $\langle O\rangle$ and $\langle f_{0}\rangle$ for the mean-field case, as can be easily seen from Eqs. (\ref{eq4}) and (\ref{eq5}). In order to clarify if the origin of crossing of the curves is a consequence of the topology or of the presence of correlations in the system (as discussed in Section 2), we performed simulations with a small modification in the sixth rule of the model (see Section 2). For this purpose, we choose at random the agent $k$ that will suffer the noise effect, which eliminates the correlations on the noise itself and keeps only the correlations due to the interactions among nearest neighbors ($i$ and $j$ in the language of Seciton 2). We have verified that the quantities $\langle O\rangle$ and $\langle f_{0}\rangle$ can also be fitted with second-order polynomials, but different from those of the previous case. In this case, the crossing of the mentioned quantities also occurs, but in a different point, $p^{**}\approx 0.82$ (not shown). Thus, we can conclude that the crossing of the $\langle O\rangle$ and $\langle f_{0}\rangle$ curves for the 2d case is a consequence of the correlations introduced by the topology on the system, i.e., by the fact that a given agent $i$ can interact (in a voter-like way) only with one of his four nearest neighbors, that are always the same during all the dynamics. The presence of correlations on the noise causes only a shift from $p^{**}$ to $p^{*}$ of the crossing point. However, one can see from Fig. \ref{fig2} (b) that in the limit of weak noise the MF equations describe well the two-dimensional results. Thus, the second-order dependency of $\langle O\rangle$ and $\langle f_{0}\rangle$ on $p$ obtained numerically can be seen as a correction of Eqs. (\ref{eq4}) and (\ref{eq5}) for the 2d case. In this case, for small values of $p$ the term $p^{2}$ can be neglected, which explains the agreement between the two approaches (2d and MF) for low $p$.

\begin{figure}[t]
\begin{center}
\vspace{0.5cm}
\includegraphics[width=0.47\textwidth,angle=0]{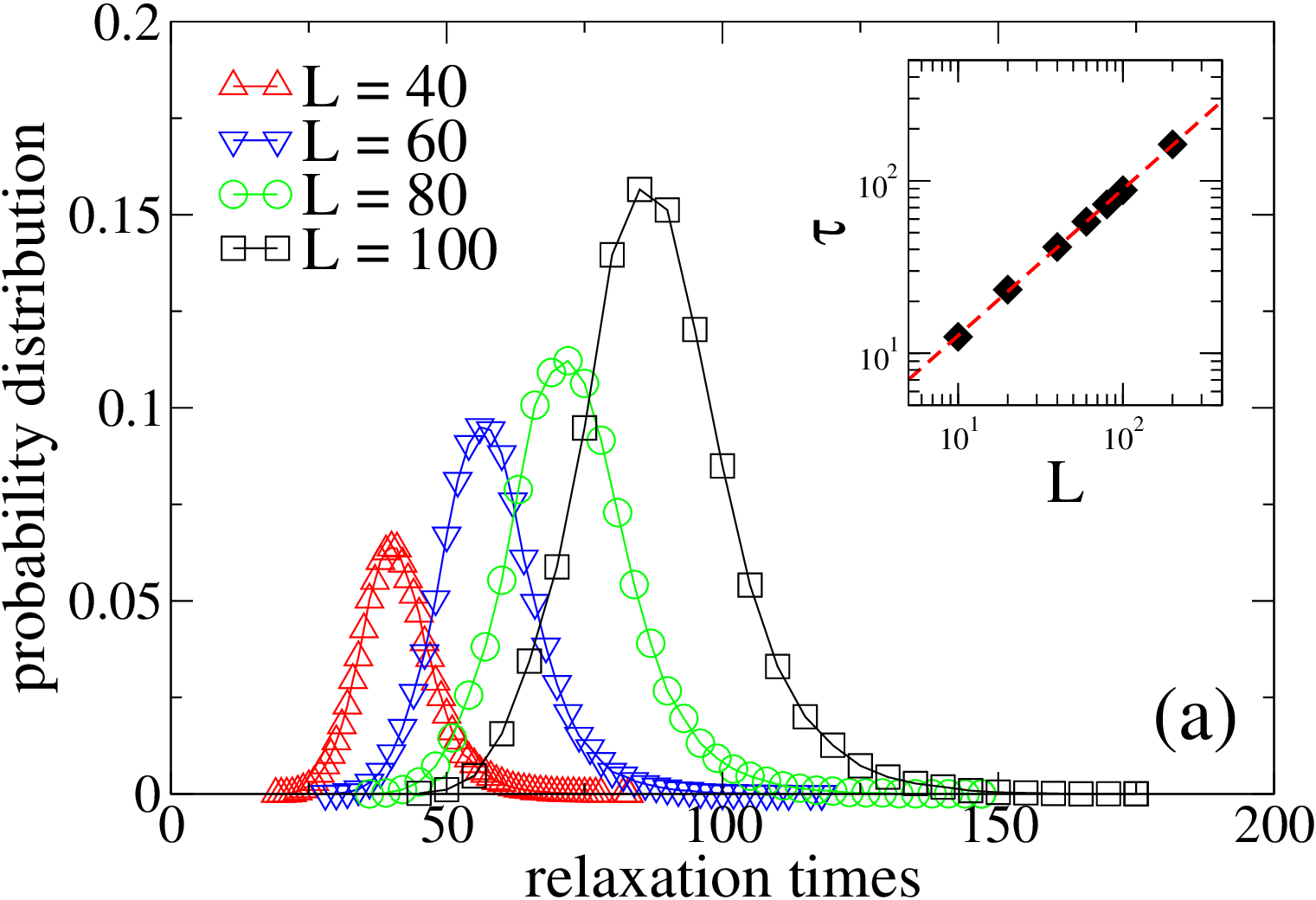}
\hspace{0.3cm}
\includegraphics[width=0.45\textwidth,angle=0]{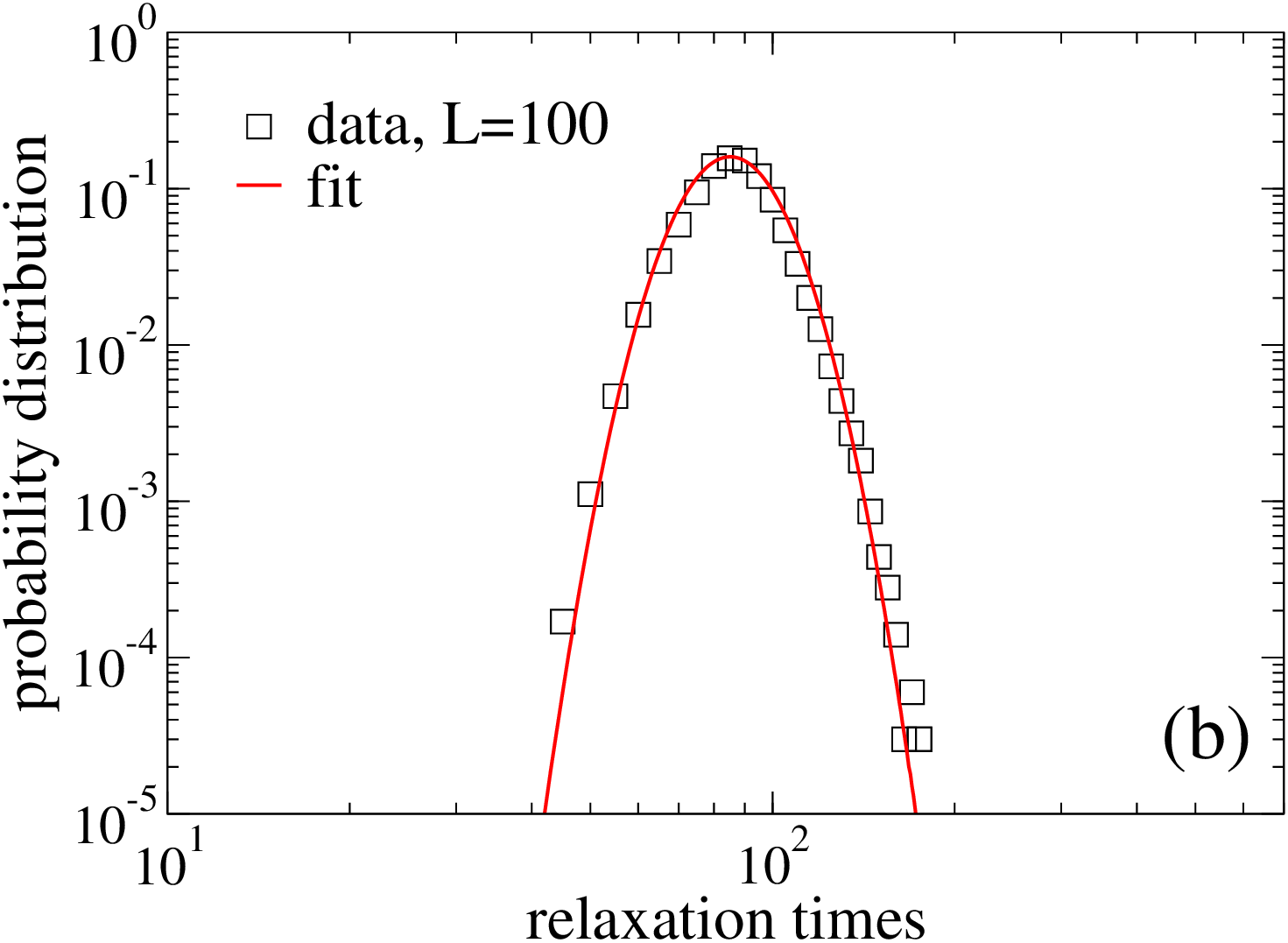}
\end{center}
\caption{Distribution of the relaxation times of the model for $p=0.0$ and different lattice sizes $L$ (a). In the inset we show the average relaxation time $\tau$  as a function of $L$ in the log-log scale. Fitting data, we obtained $\tau\sim L^{\alpha}$, where $\alpha\approx$ 0.9. It is also exhibited the relaxation times for $L=100$ in the log-log scale (b). Data are well fitted by a log-normal distribution, Eq. (\ref{eqx}).}
\label{fig3}
\end{figure}

We have also analyzed the relaxation times to consensus of the model. In this case, we have simulated $10^{5}$ samples of the model for $p=0.0$ in order to build a probability distribution of the mentioned relaxation times. In Fig. \ref{fig3} (a) we exhibit the distributions of the relaxation times for different lattice sizes $L$. As expected, when we increase $L$ the average relaxation time $\tau$ also increases. Based on the numerical data, we have found a power-law relation between $\tau$ and $L$, 
\begin{equation} 
\tau \sim L^{\alpha} ~,
\end{equation}
\noindent
where $\alpha=0.9 \pm 0.1$ (see the inset of Fig. \ref{fig3} (a)). In order to verify if the relaxation-time distribution is given by a known function, we plotted the numerical data in the log-log scale (see Fig. \ref{fig3} (b) for $L=100$). In this case, the observed parabola suggests that the times needed to find all the agents at the end having one of the extreme opinions $+1$ or $-1$ are distributed according to a log-normal function. To confirm this hypothesis we fitted the data for $L=100$ with the log-normal function 
\begin{equation} \label{eqx}
f(x)=\frac{a}{x}\,\exp[-b\,(\log\, x-x_{o})^{2}] ~.
\end{equation}
\noindent
The best fit to data is exhibited by a full line in Fig. \ref{fig3} (b), where the parameters are given by $a=13.8\pm 0.3$, $b=19.4\pm 0.2$ and $x_{o}=4.5\pm 0.1$. The function $f(x)$ is also able to fit the data for other lattice sizes $L$. It is important to mention that log-normal-like distributions of the relaxation times were also observed in other opinion models \cite{meu_physicaA,chang,schneider}. 

As a final comment, we have verified numerically that if we consider the convictions of the undecided agents as a random number in the range $[0,1]$, as was done for the agents with $s=+1$ and $s=-1$, the results are qualitatively the same. The only difference is the time needed to the system to reach the steady states. The probability distribution of the relaxation times to consensus (for $p=0.0$) is also a log-normal, and the $p$-dependency of the order parameter and of the fractions of the three opinions is the same as indicated in Fig. \ref{fig2}. In this sense, the key ingredients of the model are the presence of noise and the change in the agents' convictions, i.e., the fact that the persuaded individual assumes the conviction of the agent who influenced him.


\subsection{The model in the presence of external effects}

\quad In the presence of external effects the system follows the rules presented in section 2 but in addition there is another source of influence. This external influence acts in the system as a mass-media effect, and is modeled here in a simple way. First, we have considered that the media is favorable to opinion $+1$ \cite{meu_physicaA,babak}. In addition, the media influence is measured by a parameter $q$, that works as a probability of a certain individual to follow the media opinion. Remember that in the previous section, in the absence of the mass media, we have chosen three different agents: a random agent $i$, one of his nearest neighbors $j$ and after another different nearest neighbor, $k$. We will consider in this section that the remaining two nearest neighbors of $i$, say $x$ and $y$, that were not affected by the interaction with $i$ and by the noise $p$ are choosen to suffer the media influence. After the application of the six rules discussed in section 2, we consider that:

\begin{itemize}

\item The two mentioned nearest neighbors of $i$, $x$ and $y$, are choosen;

\item If $x$ has opinion $s_{x}=-1$ or $s_{x}=0$, he will follow the media opinion ($+1$) if his conviction is lower than the media influence, i.e., if $C_{x}<q$ we update $s_{x}=+1$. In this case, we generate another conviction for agent $x$: a real number obtained from a uniform distribution $[0,1]$;

\item The same rule is applied to $y$: if $y$ has opinion $s_{y}=-1$ or $s_{y}=0$, he will follow the media opinion ($+1$) if his conviction is lower than the media influence, i.e., if $C_{y}<q$ we update $s_{y}=+1$. In this case, we generate another conviction for agent $y$: a real number obtained from a uniform distribution $[0,1]$;

\end{itemize}
\begin{figure}[t]
\begin{center}
\vspace{0.5cm}
\includegraphics[width=0.4\textwidth,angle=0]{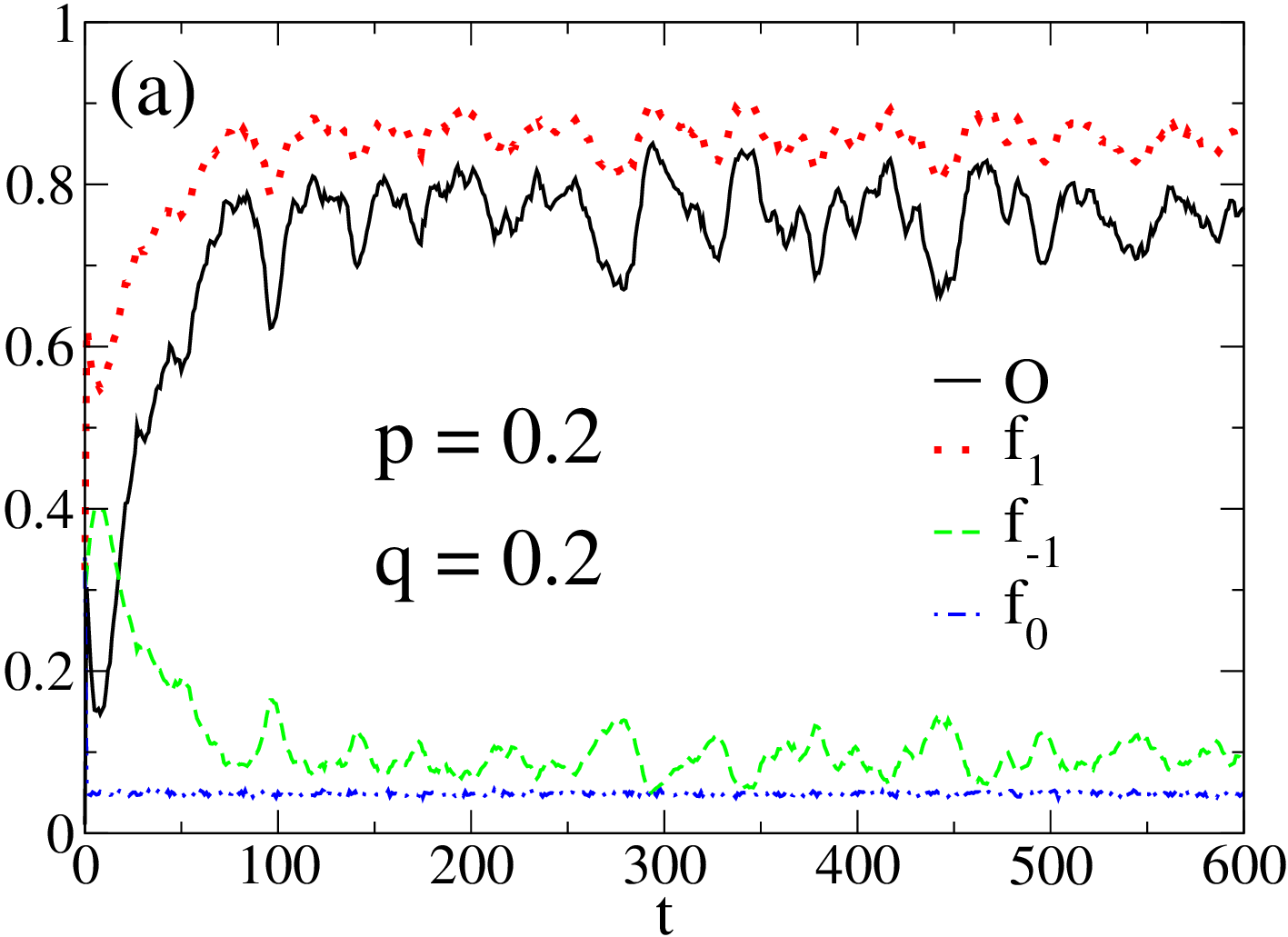}
\hspace{0.5cm}
\includegraphics[width=0.4\textwidth,angle=0]{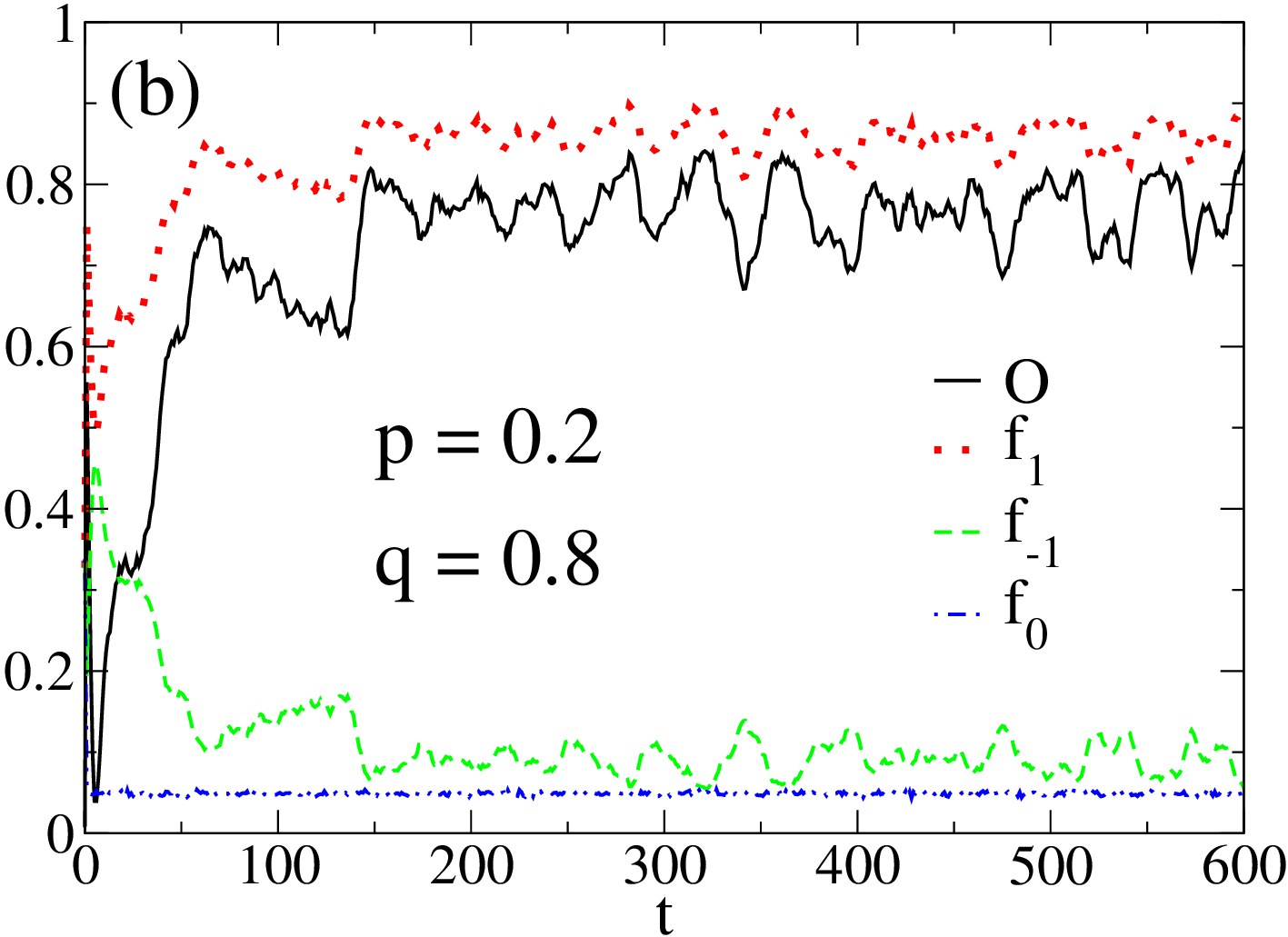}
\\
\vspace{1.0cm}
\includegraphics[width=0.4\textwidth,angle=0]{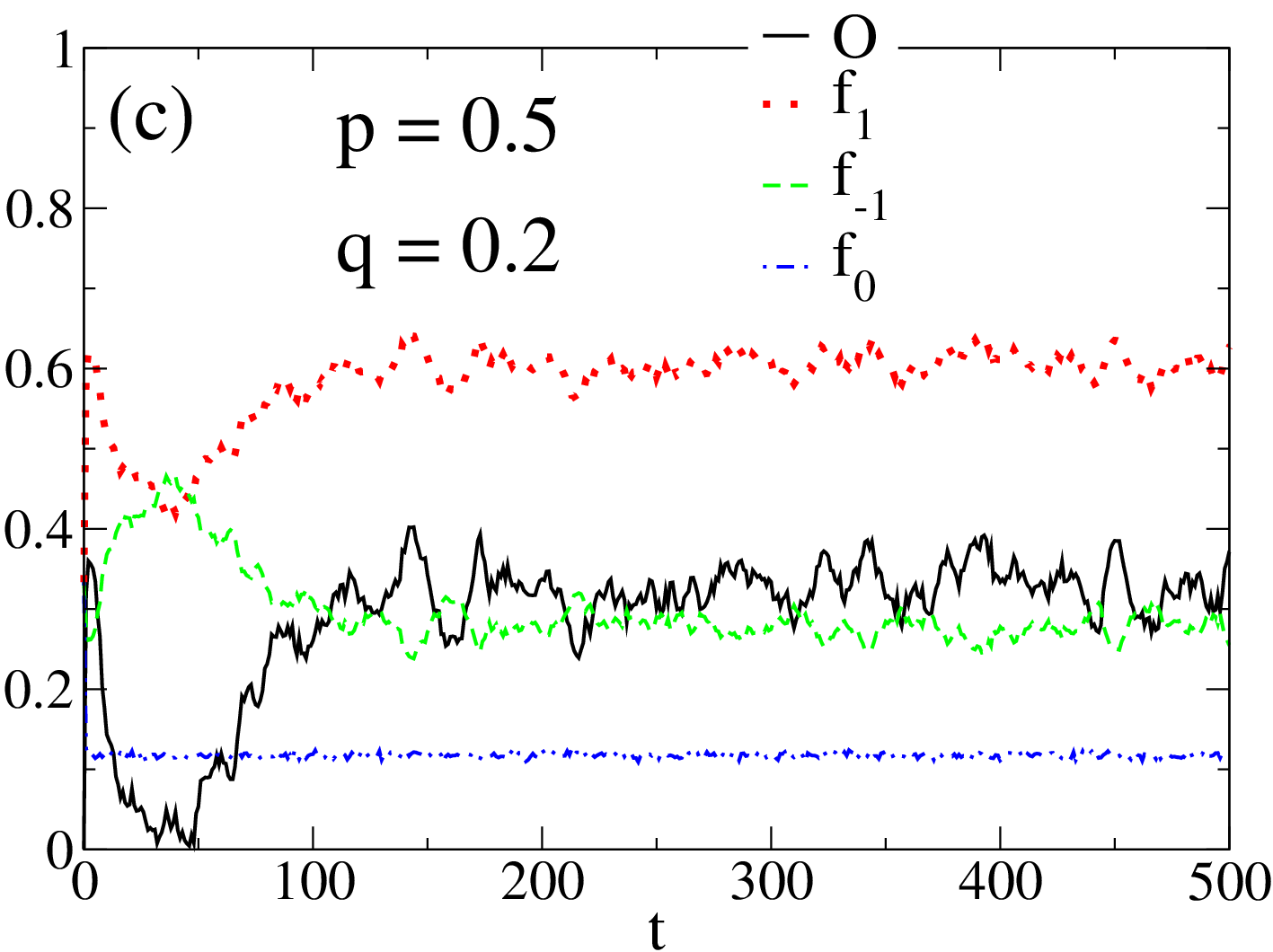}
\hspace{0.5cm}
\includegraphics[width=0.4\textwidth,angle=0]{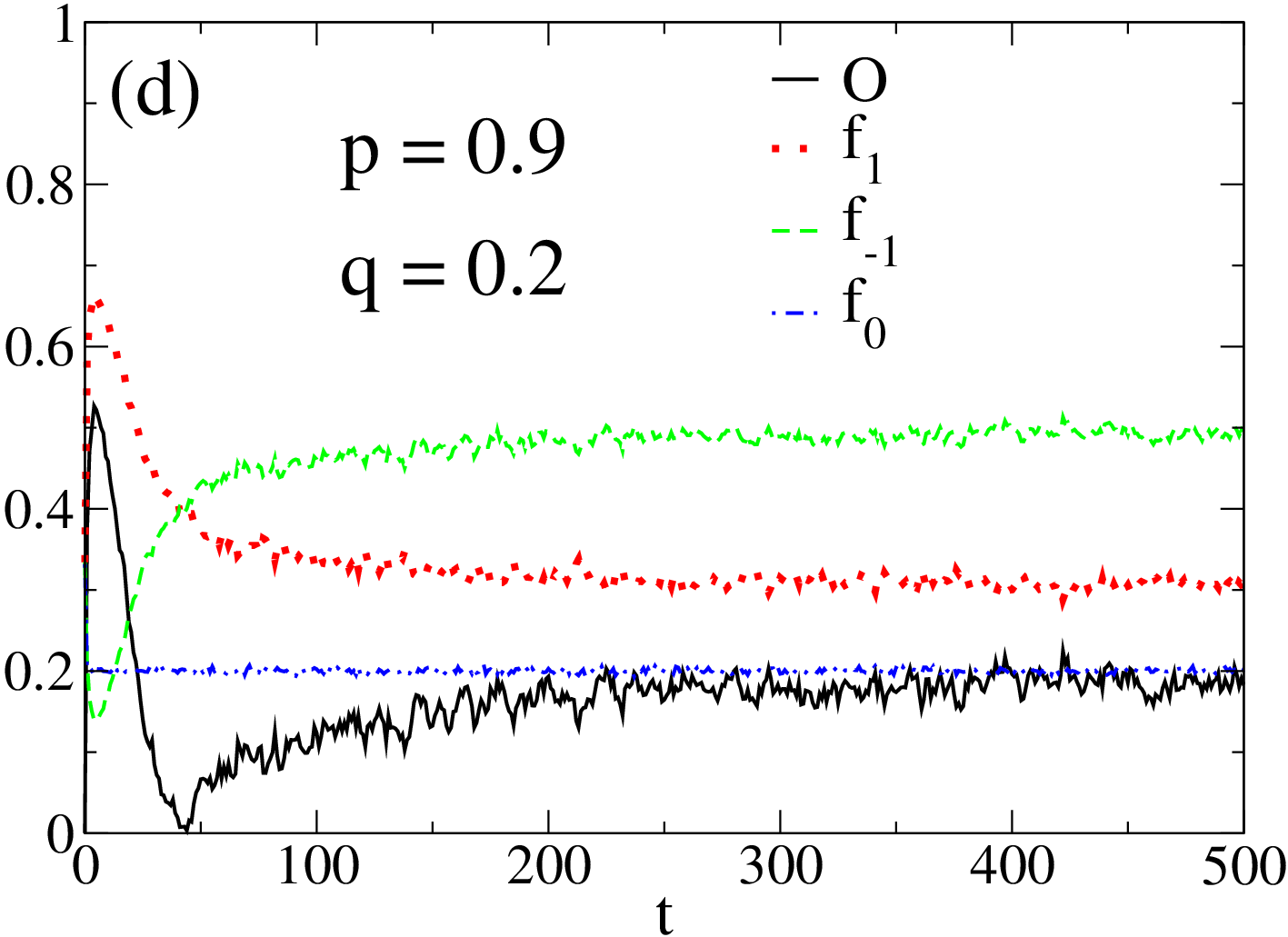}
\end{center}
\caption{Time evolution of the order parameter $O$ and the fractions $f_{1}$, $f_{-1}$ and $f_{0}$ of the agents with opinions $+1$, $-1$ and $0$, respectively, in the presence of mass media. Results are for lattice size $L=100$ and parameters $p=0.2$ and $q=0.2$ (a), $p=0.2$ and $q=0.8$ (b), $p=0.5$ and $q=0.2$ (c) and $p=0.9$ and $q=0.2$ (d).
}
\label{fig4}
\end{figure}

Thus, if the agent $x$ ($y$) has a conviction $C_{x}$ ($C_{y}$) greater than the media influence $q$, he will not follow the media opinion. Differently from the case of a given agent, it is difficult to quantify the media conviction. Thus, if a certain individual is persuaded by the media to change his opinion, we generate a conviction for him from a uniform distribution $[0,1]$. In Fig. \ref{fig4} we exhibit the time evolution of the order parameter $O$ and of the fractions of each opinion for typical values of the parameters $p$ and $q$. We can see that the system reaches steady states for all values of the parameters as in the case with no external effects. In Figs. \ref{fig4} (a) and (b) we fixed the value of the noise $p=0.2$ and compare the effects of varying the mass-media parameter $q$. In this case,  one can see that the noise parameter dominates the behavior of the system, and the fractions $f_{1}$, $f_{-1}$ and $f_{0}$ after a long time are similar for the two considered media parameters, namely $q=0.2$ and $q=0.8$. On the other hand, if we compare the evolution of the system for a fixed value of the media parameter ($q=0.2$) and different values of $p$, one can observe a distinct behavior of the system [see Figs. \ref{fig4} (c) and (d)]. In fact, in this case the fractions $f_{1}$, $f_{-1}$ and $f_{0}$ after many time steps are different for the two considered values of $p$, namely $p=0.5$ and $p=0.9$.

\begin{figure}[t]
\begin{center}
\vspace{0.5cm}
\includegraphics[width=0.47\textwidth,angle=0]{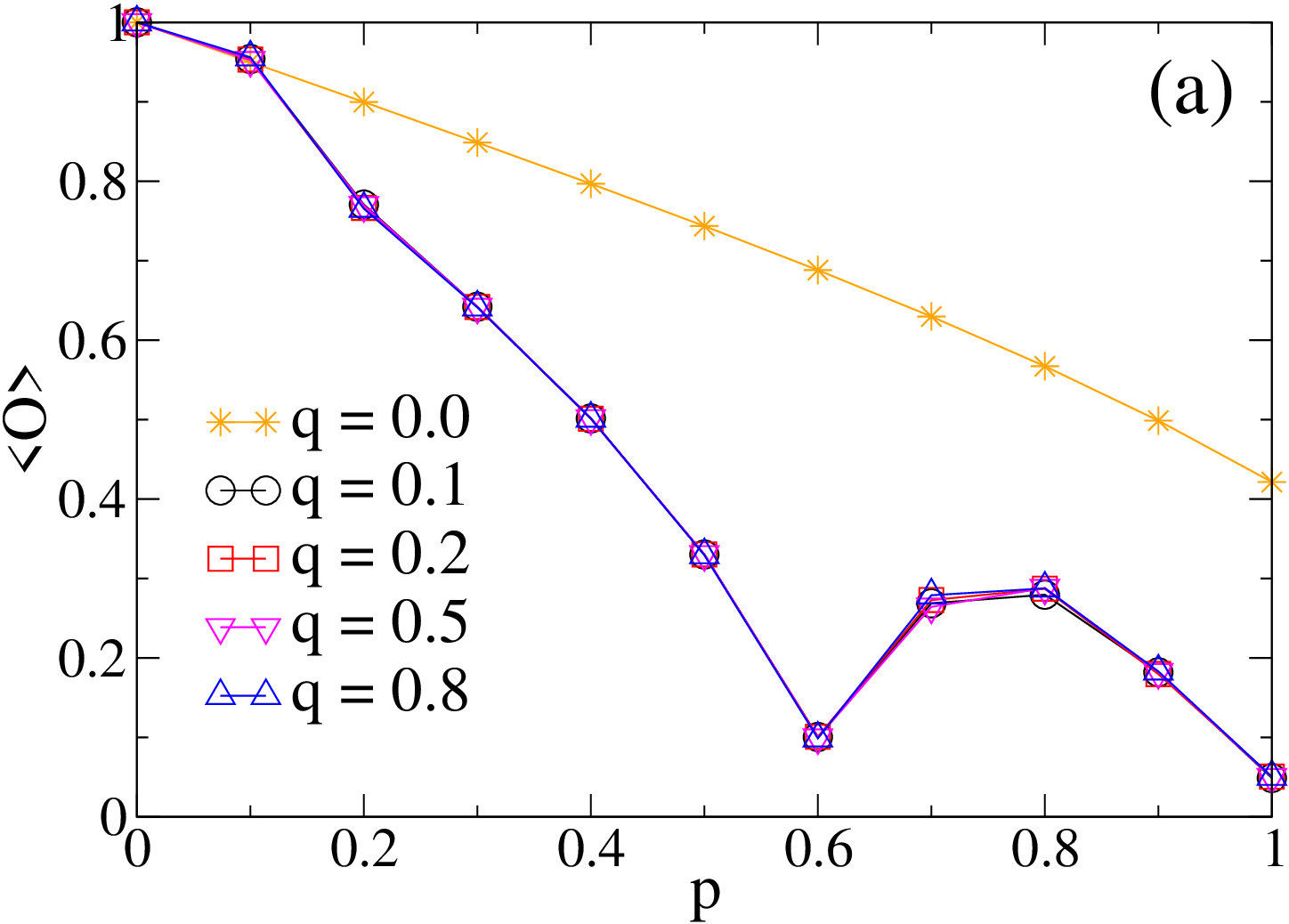}
\hspace{0.3cm}
\includegraphics[width=0.46\textwidth,angle=0]{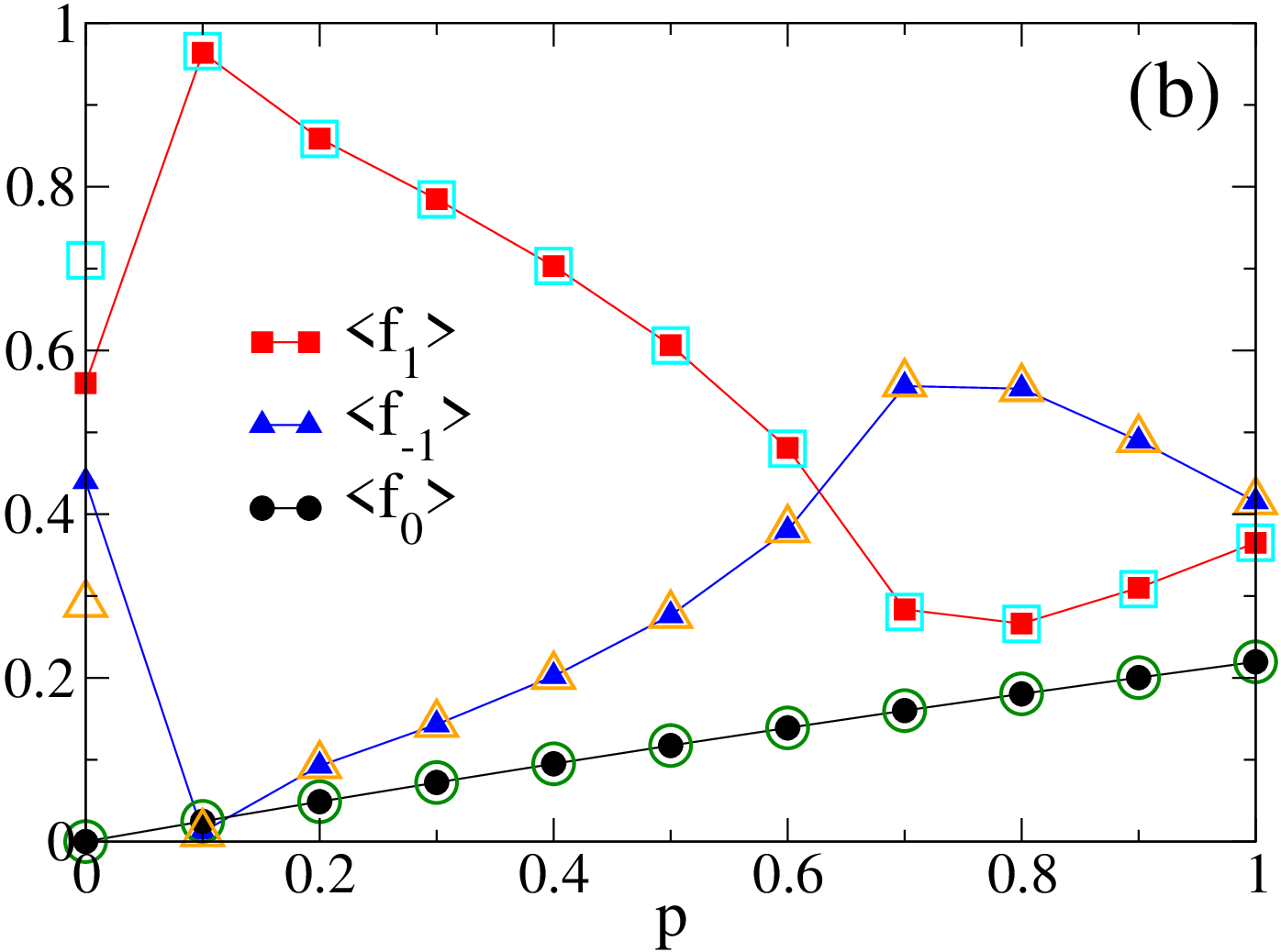}
\end{center}
\caption{Average stationary order parameter $\langle O\rangle$ as a function of $p$ for typical values of the media parameter $q$ (a). It is also shown the average stationary fractions of the three opinions $\langle f_{1}\rangle$, $\langle f_{-1}\rangle$ and $\langle f_{0}\rangle$ versus $p$ (b) for $q=0.2$ (full symbols) and $q=0.8$ (empty symbols). 
}
\label{fig5}
\end{figure}

In Fig. \ref{fig5} (a) we exhibit the average stationary order parameter $\langle O\rangle$ as a function of $p$ for typical values of the media parameter $q$. We can see that for small values of $p$ like $p=0.1$ the consequences of the external effects are negligible, but for $p>0.1$ the media affects strongly the system. In this case, one can observe that $\langle O\rangle$ decreases fast until $p=0.6$ for all values of $q$. On the other hand, the value of the order parameter increases in the range $0.6 < p < 0.8$, and decreases again for $p>0.8$. The cause of this effect will be clear in the following.

In Fig. \ref{fig5} (b) we exhibit the average stationary fractions of opinions $+1$, $-1$ and $0$ ($\langle f_{1}\rangle$, $\langle f_{-1}\rangle$ and $\langle f_{0}\rangle$, respectively) for two typical values of the mass-media parameter $q$, namely $q=0.2$ (full symbols) and $q=0.8$ (empty symbols). One can see that the fractions of extreme opinions $\langle f_{1}\rangle$ and $\langle f_{-1}\rangle$ are strongly affected by the noise $p$ for a fixed value of the mass-media parameter $q$, as was discussed above (see Fig. \ref{fig4}). The fraction on undecided individuals $\langle f_{0}\rangle$ grows linearly with $p$, i.e., in a slower way in comparison with the case with no media [see Fig. \ref{fig2} (b)]. On the other hand, for a fixed $p$ there are no differences between the two exhibited values of $q$, except for the noiseless case $p=0.0$ (see also Fig. \ref{fig4}). In the last case, the fraction of $+1$ ($-1$) opinions increases (decreases) when we increase the mass-media influence $q$, as expected, since the media persuades the agents with opinions $-1$ and $0$ to change to $+1$, and the individuals do not come back to the undecided state because $p=0.0$. On the other hand, the fraction of $s=0$ opinions for $p=0.0$ is not affected by the variation of the parameter $q$.

As the media favors opinion $+1$, it could be expected that this opinion will be the majority opinion in the population at the steady states. However, this behavior is observed only for $p<\sim 0.6$, for all values of $q$ in the range $0.0 < q < 1.0$. In Fig. \ref{fig5} (b) we exhibit two distinct cases, namely $q=0.2$ and $q=0.8$, to illustrate this result. On the other hand, for $p>\sim 0.6$ the opinion contrary to the media opinion, i.e. $-1$, dominates the majority of the agents in the steady states [see Figs. \ref{fig4} (d) and \ref{fig5} (b)]. In analogy with elections, this means that if there are many agents becoming undecided, the media candidate (related to opinion $+1$) will not win the election, even for a strong effect of the propaganda. This fact explains the effect observed in the order parameter for $0.6 < p < 1.0$ [see Fig. \ref{fig5} (a)]. In fact, as one can write the stationary order parameter as $\langle O\rangle=|\langle f_{1}\rangle - \langle f_{-1}\rangle|$ and $\langle f_{-1}\rangle > \langle f_{1}\rangle$ for $p>\sim 0.6$, the value of $\langle O\rangle$ in this case is greater than in the case $p=0.6$ since in this last case we have $\langle f_{-1}\rangle \approx \langle f_{1}\rangle$.

\begin{figure}[t]
\begin{center}
\vspace{0.5cm}
\includegraphics[width=0.52\textwidth,angle=0]{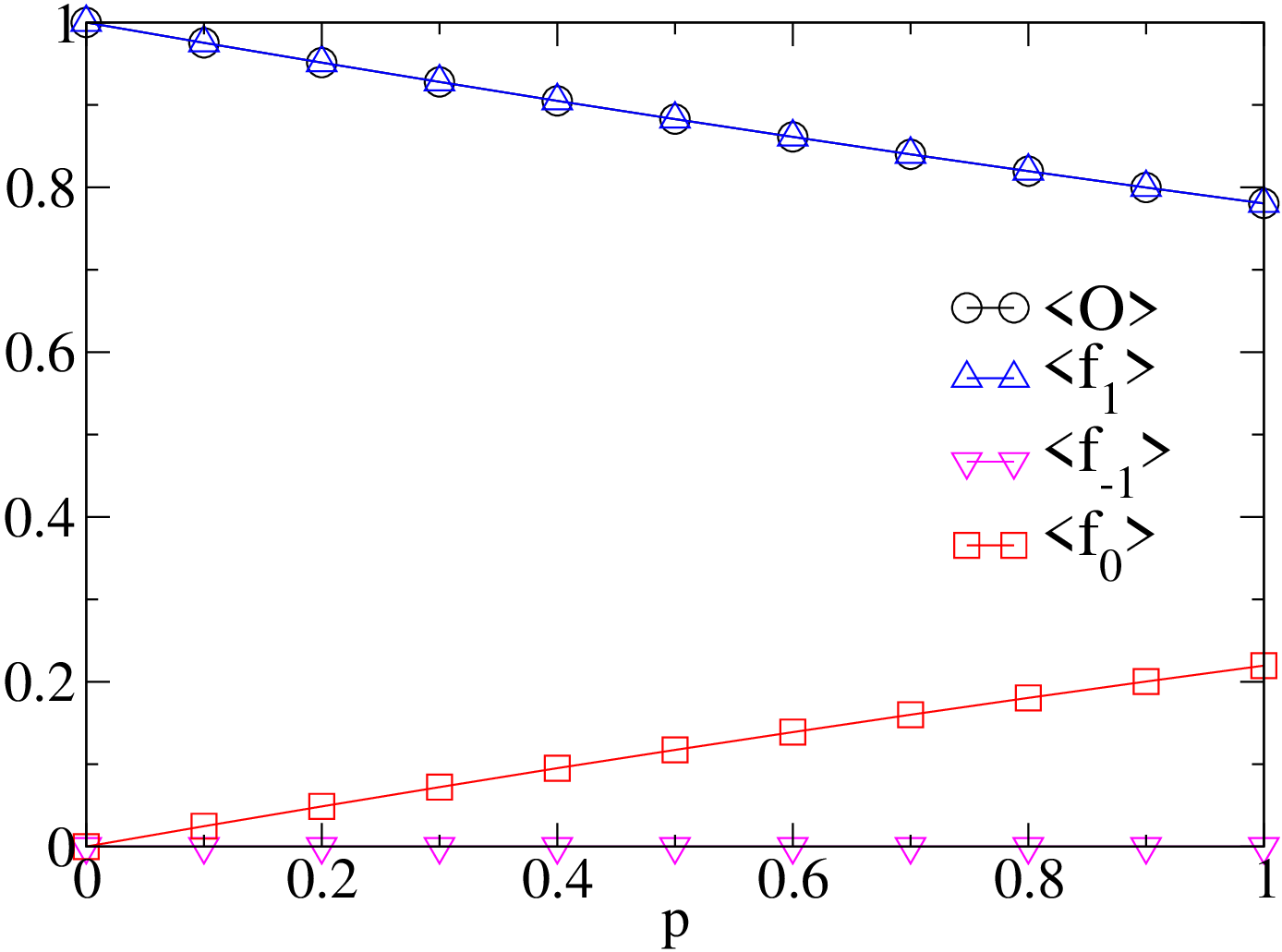}
\end{center}
\caption{Stationary fractions of the order parameter and of the three opinions versus $p$ for $q=1.0$.
}
\label{fig6}
\end{figure}

In Fig. \ref{fig6} we exhibit the average stationary values of the order parameter and of the fractions of opinions $+1$, $-1$ and $0$ for $q=1.0$. As in the case with no mass-media effects, the consensus states are only possible in the absence of noise ($p=0.0$), even if the media effect always persuade the agents ($q=1.0$). In other words, even a sufficient small noise $p$ leads each agent in the population to be in one of two different states, $s=+1$ or $s=0$, with the opinion contrary to the media opinion ($-1$) being extincted from the system, i.e., we have $\langle f_{-1}\rangle=0$ for all values of $p$ (see Fig. \ref{fig6}). As a consequence, we have $\langle O\rangle = \langle f_{1}\rangle$, and we observe linear behaviors of $\langle f_{1}\rangle$ and $\langle f_{0}\rangle$ with $p$: while $\langle f_{0}\rangle$ grows linearly with the increase of the noise $p$, $\langle f_{1}\rangle$ decays linearly with $p$.

\section{Final Remarks}

\qquad In this work, we have studied a voter-like model on square lattices with linear sizes $L$. The individuals may be in one of three distinct states, represented by opinions $s=+1$, $-1$ or $0$. The agents change their opinions via two competing mechanisms. First, two agents may interact via a voter-model dynamics, but the usual copying process is limited by the agents' convictions. Furthermore, we have considered that in a given interaction the persuaded agent takes the opinion and the conviction of the agent that interacted with him. To our knowledge, the presence of convictions was analyzed in some works (see for example \cite{ben-naim1,ben-naim2,ben-naim3,sen,souza,meu_celia}), but not as we done. The second mechanism is a noise $p$, which allows agents to change their opinions to the undecided state $s=0$. As the agents that participate of the dynamics in a given time step are nearest neighbors in the square lattice, the dynamics inserts correlation effects on the system and on the noise itself. These rules are realistic ingredients to be considered in opinion dynamics.

We have verified that the system reaches consensus only in the absence of noise ($p=0.0$), and in this case the relaxation times are log-normally distributed. The average value of the distribution grows with the system size as $\tau \sim L^{\alpha}$, where $\alpha\approx 0.9$. For $p>0.0$ we have observed that one of the extreme opinions $+1$ or $-1$ disappears of the system, and the surviving extreme opinion coexists with the $0$ opinion. The fraction of $s=0$ opinions in the stationary states grows fast with increasing values of $p$, and a numerical analysis of the simulation data suggests that the above-mentioned surviving opinion is the majority opinion in the population for $p < p^{*}$, being overcome by the opinion $0$ for  $p > p^{*}$, where $p^{*}\approx 0.9$.

We extended the model to consider external effects. These effects act in the system as a mass medium (TV, radio, ...) and are quantified by a parameter $q$. If $q$ is greater than the conviction of a given agent, this agent follows the media opinion, that we have considered as $+1$. The first consequence of the media effect appears in the stationary fraction of $s=0$ opinions, that grows linearly with the noise $p$ for all values of $q>0$, i.e., in a slower way than in the absence of external effects. In the absence of noise ($p=0.0$), the scenario is similar to the case with no media effect, i.e., the system reaches consensus but in a faster way than in the case $q=0.0$, but the $-1$ opinion does not disappears of the system. In the other extreme situation, $q=1.0$, the opinion $-1$ disappears of the system in the stationary states for all values of the noise $p$, and the fraction of $+1$ opinions decays linearly with the increase of $p$. For the intermediary values of $q$, i.e., in the range $0.0 < q < 1.0$, we have observed that the opinion $-1$ persists in the long-time limit for all values of the noise $p$. Based on the simulation data we verified that, for a given value of $q$, the media opinion $+1$ in the stationary states is the majority opinion in the population only for $p < \sim 0.6$, being overcome by the opinion $-1$ for  $p > \sim 0.6$. In other words, even a strong effect of advertising may not be enough to make winner the opinion supported by the mass media, which is a nontrivial result.

The real dynamics of the human societies is very complicated. However, our model can simulate real situations in communities. In any public debate, the discrete opinions considered in our model can represent extreme favorable opinion ($s=+1$), extreme unfavourable opinion ($s=-1$) or indecision ($s=0$). Thus the order parameter in the model corresponds to the overall rating and the ordered state means there is a clear decision made. For example, in an electoral process with two different candidates A and B, each elector (agent) votes for the candidate A (opinion $s=+1$), for the candidate B (opinion $s=-1$) or is undecided (opinion $s=0$). The competition of the convictions in an interaction can be seen as the competition between the persuasiveness of an agent and the resistance of the other agent to change your candidate. The noise $p$ represents the volatility of some individuals, that tend to spontaneously change your choices. Finally, the external effect introduced in Section 3.2 models the effect of advertising on the individuals' choices. In fact, the electorate is susceptible to the effects of electoral surveys broadcasted by the mass media. Furthermore, the occurrence of consensus states with $O=1$ are usually related to dictatorships, whereas the states with $O<1$ are desirable since they represent democracy-like situations \cite{schneider,meu_bjp}.

Another example is the dynamics of agents on financial markets, where the opinions represent the decision for selling stocks ($s=+1$), the decision for buying stocks ($s=+1$) or the decision to do nothing ($s=0$). In this case, when a pair of agents with opposite decisions $+1$ and $-1$ interacts, one of the two agents can be persuaded to change your choice (the agent with lower conviction). The noise $p$ in this case represents the decision of an agent to stop their actions of selling or buying stocks, maybe due to fluctuations on the stock prices. On the other hand, the external effect models the campaign made by some companies (in television, for example) in order to sell their stocks.

The presence of special individuals like contrarians, inflexibles, opportunists or stubborn agents was considered in some opinion models \cite{mobilia,babak,schneider,galam_inflex}. It would be interesting to extend the present model to include the above-mentioned agents and study the impact of their actions on opinion formation. In addition, one can also study the properties of the model in various lattices and networks.


\section*{Acknowledgments}

The author acknowledges financial support from the Brazilian funding agencies FAPERJ and CAPES. We also acknowledge thoughtful remarks by anonymous referees which significantly improved the text.

\end{document}